\documentclass[12pt]{article}
\usepackage{group21}
\usepackage{epsfig}
\title{ Tiles--inflation rules for the class of canonical 
tilings ${{\bf\cal T}}^{\bf *(2F)}$   
derived by the projection method}
\author{Zorka Papadopolos, Christoph Hohneker and Peter Kramer }
\address{Institut f\"ur Theoretische Physik,
         Universit\"at T\"ubingen,\\
         Auf der Morgenstelle 14, 72076 T\"ubingen, Germany}
\def\ts2f{${\cal T}^{*(2F)}$}
\def\tsa4{${\cal T}^{*(A_4)}$}
\def\tms{${\cal T}^{(MS)}$}
\def\tp{${\cal T}^{(P)}$}
\newcommand{\ZZ}{{{\mbox{Z}}\!\!{\mbox{\bf Z}}}}

\newcommand{\EE}{{{\mbox{I}}\!{\mbox{\bf E}}}}
\newcommand{\pp}{\pi_{\parallel}}
\newcommand{\ps}{\pi_{\perp}}
\newcommand{\Az}{{{\mbox{$\bigcirc$}}\!\!\!\!\!\!\:{\mbox{2}}\,}} 
\newcommand{\Af}{{{\mbox{$\bigcirc$}}\!\!\!\!\!\!\:{\mbox{5}}\,}} 

\begin{document}
\maketitle

\centerline{ Dedicated to Ludwig  Danzer on the occasion of 
             his 70th birthday }

\begin{abstract}

The tiles of the canonical  tilings \ts2f \ are six tetrahedra. 
We determine their inflation rules by the projection method.

\end{abstract}

\section{Introduction}

\subsection{2F--module}
The standard icosahedral projection is defined by the particular 
embedding of the icosahedral group $Y\cong A(5)$ into the 
hyperoctahedral group $\Omega(6)=(\ZZ/2\ZZ)^6 \times_s S(6).$
Upon subducing the defining representation 
$D$ of $\Omega(6)$ to $A(5)$, 
the reduction to irreducible form is obtained as
$BD(g)B^{-1}=D^{[31^2_+]}(g)+D^{[31^2_-]}(g)$,
where $B$ is chosen

\begin{equation}
B=\sqrt{\frac{1}{2(\tau +2)}} \left[
\begin{array}{cccccc} 
0 & 1 & \bar{1} & \bar{\tau} & 0 & \tau \\
1 & \tau & \tau & 0 & \bar{1} & 0 \\
\tau & 0 & 0 & 1 & \tau & 1 \\
0 & \tau & \bar{\tau} & 1 & 0 & \bar{1} \\
\tau & \bar{1} & \bar{1} & 0 & \bar{\tau} & 0 \\
\bar{1} & 0 & 0 & \tau & \bar{1} & \tau
\end{array}  
\right],
\end{equation}
and $\tau $ is the golden ratio, $\tau = ( 1 + \sqrt {5} )/2 $.
The symbol $\bar x$ stands for $-x$. 
The columns of this matrix form six orthogonal basis vectors 
of $\ZZ^6$.
The first three rows of the matrix $B$ give the six basis 
vectors of $\ZZ^6$, $ \{ e_i|(e_i,e_j)={\delta }_{ij} \} $, 
``icosahedrally projected'' 
to the ``parallel space'', 
$e_{i\parallel}\in \EE_\parallel$ (3D space of the irreducible 
representation $D^{[31^2_+]}(g)$), and the next three 
rows give the six vectors ``icosahedrally projected'' to the 
``orthogonal space'' (3D space of the irreducible representation 
$D^{[31^2_-]}(g)$), $e_{i\perp}\in \EE_\perp $. The Weyl group 
of the lattice $D_6$, $W_{D_6}$ is a subgroup of the 
hyperoctahedral group,
$W_{D_6}=(\ZZ/2\ZZ)^5 \times_s S(6)<\Omega(6)$.
The icosahedrally projected points of the 
$D_6\equiv 2F$--lattice into $\EE_\parallel$ we call 
$2F$--module. 

It can be shown that the $GL(6,\ZZ)$ transformation of $D_6$
points such that it commutes with  $A(5)$
reduced to $\EE_\parallel + \EE_\perp$
and acts in $\EE_\parallel$ as a minimal stretching is

\begin{equation}
I_{D_6} = \frac{1}{2} \left[
\begin{array}{cccccc}  
{\mbox {$1$}}&{\mbox {$1$}}&{\mbox {$1$}}&{\mbox {$1$}} & 
{\mbox {$1$}} &  {\mbox {$1$}} \\
{\mbox {$1$}}&{\mbox {$1$}}&{\mbox {$1$}}&{\mbox{$ \bar {1}$}}&
{\mbox {$\bar {1}$}}&{\mbox {$1$}}\\
{\mbox {$1$}}&{\mbox {$1$}}&{\mbox {$1$}}&{\mbox {$1$}}&
{\mbox{$\bar {1}$}}&{\mbox {$\bar {1}$}}\\
{\mbox {$1$}}&{\mbox {$\bar {1}$}}&{\mbox {$1$}}&{\mbox {$1$}}&
{\mbox {$1$}}&{\mbox {$\bar {1}$}}\\
{\mbox {$1$}}&{\mbox {$\bar {1}$}}&{\mbox{$ \bar {1}$}}&
{\mbox {$1$}}&{\mbox {$1$}}&{\mbox {$1$}}\\
{\mbox {$1$}}&{\mbox {$1$}}&{\mbox{$ \bar {1}$}}&
{\mbox {$ \bar {1}$}}&{\mbox {$1$}}&{\mbox {$1$}}
\end{array}  
\right].
\end{equation}
This matrix is written in the ${\ZZ}^6$ basis. When referred to
a $D_6$ basis it becomes an element of $GL(6,\ZZ)$.
The minimal stretching in Ogawa's sense,  such that the
quasilattice points are mapped into quasilattice points,
is a factor $\tau$, and $I_{D_6}$ transformed to 
$\EE_\parallel + \EE_\perp$ is

\begin{equation}
\tau\pp   -\frac{1}{\tau}\ps,
\end{equation}
where $\pp$ is a projector to $\EE_\parallel$ and 
$\ps$ to $\EE_\perp$.
The basis of the 2F--module can be chosen as for example

\begin{equation}
B^{D_6}=\{ e_{3\parallel } + e_{6\parallel }, 
e_{2\parallel } + e_{5\parallel }, e_{1\parallel } - 
e_{4\parallel } \}.
\end{equation}
The coefficients in each of the three directions are 
in $\ZZ [\tau ] $, so it is a $\ZZ [\tau ]$--module with 
inflation factor $\tau $ (the smallest stretching factor 
by which module points are mapped into module points).

\subsection{Canonical tiling \ts2f }

The acceptance domain, i.e. the vertex--window $W$ of the 
tilings 
\ts2f \cite{A}
is the Voronoi 
domain ($V$) of the root--lattice $D_6$ icosahedrally 
projected to $\EE_\perp $, $V_\perp = W $. In $\EE_\perp $, its 
outer shape is a triacontahedron with the edge 
length $\Af$ (parallel to the $5$--fold symmetry axes of an 
icosahedron), $\Af = \frac{1}{\sqrt {2}} $.
The vertices of the Voronoi domain of the root--lattice 
$D_6$ are the
three translationally nonequivalent ``holes"\cite{CS} of the 
$D_6$ 
lattice.  We denote them \cite{A} by $a$, $b$ and $c$. After the
projection of the Voronoi domain to $\EE_\perp $, the vertices
of the window $V_\perp = W$ are of  type $a$ and $c$, marked 
by black and
white balls (Figure 1) respectively. The vertices of the tiles
(prototiles) in the class of tilings \ts2f \ are the  $D_6$
root lattice points projected to $\EE_\parallel$.  
The tilings are obtained by the ``cut and project'' method: 
cut the 3D boundary $X(3)$ of $V$, project its dual 
boundary $X^*(3)$ to $\EE_\parallel $. $X^*_\parallel $ is 
coded by $X(3)$ projected to $\EE_\perp $, $X_\perp $.
All codings are the projected boundaries of $V$ 
(that we together with $V$ call the Voronoi complex) 
to  $\EE_\perp $.
Dual boundaries are the 3D boundaries of Delaunay cells.
In general: all tiles, faces, edges, vertices, vertex 
configurations from the tilings in $\EE_\parallel $ are 
coded by corresponding dual objects in $\EE_\perp$, i.e. by the
Voronoi complex projected to $\EE_\perp$. 
Consequently, 
all properties of the tilings  
in $\EE_\parallel $ can be determined in $\EE_\perp $.
This ``projection species" \ts2f are equivalent \cite{Ko} 
to De Bruijn's strip projection species of Danzer \cite{Da}.

The prototiles of the class of nonsingular locally
isomorphic  tilings \ts2f \  are six 
tetrahedra with the 
edges parallel to the directions of 2--fold symmetry axes of an 
icosahedron. They have two lengths, the standard 
$\Az=\sqrt{\frac{2}{\tau + 2}}$ and $\tau\Az$.
The tiles are denoted \cite{A} by $X^*_\parallel(3)=
A^*_\parallel, B^*_\parallel, C^*_\parallel, D^*_\parallel,
F^*_\parallel $ and $ G^*_\parallel$.
They are coded in $\EE_\perp $ by corresponding dual boundaries
$X_\perp(3)=A_\perp, B_\perp, C_\perp, D_\perp, F_\perp$  and  
$G_\perp$, respectively.
In Figure 1 we denote the tiles by $X^*$. 
In the paper we are not going to 
mention explicitly $X_\perp(3)$ any more, so we simplify the 
notation, instead of $X^*_\parallel $ we write $X$.

%
%
\begin{figure}[]
\vfill 
\caption{ (trixf.gif, ts2fxf.gif) 
The window $W = V_\perp $ for the class of tiling 
${\cal T}^{*(2F)}$ in $\EE_\perp $ (upper part) 
and the tiles in $\EE_\parallel $ (lower part).}
\end{figure}

The tiles (prototiles) $X= A, B, C, D, F$ and $G$ have four 
kinds of 
faces ${\Sigma }_{i  \parallel}^*$, $i=1,...,4$
that are coded in $\EE_\perp $ by the corresponding (dual)
4--boundaries of the Voronoi domain of the $D_6$--lattice,
${\Sigma }_{i\perp}$, $i=1,...,4$\cite{B} . 
All faces  ${\Sigma }_{i  \parallel}^*$ of tetrahedra are 
clearly triangles, from now on denoted by ${\Sigma }_i$.
${\Sigma }_4$ is an equilateral triangle with edge length $\Az$,
${\Sigma }_1$ is also equilateral  with edge length $\tau \Az$.  
These faces are perpendicular to the directions of 3--fold 
symmetry axes of an icosahedron. 
${\Sigma }_2$ and ${\Sigma }_3$ are the golden triangles:
${\Sigma }_2$ with one edge $ \tau \Az$ and 
two $\Az$, ${\Sigma }_3$  with two edges $\tau\Az$ and 
one $\Az$.  These faces are perpendicular to the directions of 
5--fold symmetry axes of an icosahedron. The face and edge 
content of the tiles is presented by the table

\begin{equation}
\begin{tabular}{c||cccccc} 
  & ${\Sigma }_1$ & ${\Sigma }_2$ & ${\Sigma }_3$ & ${\Sigma }_4$ & $\Az$ & $\tau\Az$ \\ 
\hline \hline
A & 2 & 0 & 2 & 0 & 1 & 5 \\ 
B & 0 & 2 & 0 & 2 & 5 & 1 \\ 
C & 1 & 1 & 2 & 0 & 2 & 4 \\ 
D & 0 & 2 & 1 & 1 & 4 & 2 \\ 
F & 0 & 0 & 3 & 1 & 3 & 3 \\ 
G & 1 & 3 & 0 & 0 & 3 & 3. \\  
\end{tabular}
\end{equation}

Let us mention three important characteristics of the tilings
\ts2f.

The tilings \ts2f  \ are dissectable into the 
tilings \tsa4 \cite{A}. 
The tilings \tsa4 \ are  particular 2D triangular tilings 
of a plane by two golden triangles ${\Sigma }_2$ and 
${\Sigma }_3$ \cite{C}.
To a fixed 5--fold direction of the triacontahedron (the
vertex--window for the \ts2f \ tilings, $W=V_{\perp}$) there is
a corresponding decagonal belt of  height 
$x=\frac{2}{\tau + 2} \Af \: $, see  Figure 2.  
%
%
\begin{figure}[h]
\caption{(belt2.gif) 
The decagonal belt of the hight $x$ orthogonal to
a 5--fold direction.}
\end{figure}
Let a plane orthogonal to the chosen 5--fold direction 
intersect the belt. 
This intersection is a decagon, the vertex window for 
the tilings \tsa4. 
The same plane intersects with a set of
corresponding\footnote {The corresponding 5-- and
4--boundaries of $V$ projected to $\EE_\perp$ are such 
that in the 3D tilings they  code
the edges and the 2D faces, all orthogonal to the chosen
5--fold direction.}  5--boundaries ($\Omega_{1\perp}$, 
$\Omega_{2\perp}$) and 4--boundaries ($\Sigma_{2\perp}$, 
$\Sigma_{3\perp}$) of  $V_{\perp}$.
These intersections form the coding of the edges and tiles of
the tilings \tsa4.
As a consequence, in $\EE_\parallel$ there appear the 
planar (2D) tilings \tsa4 \ 
which tiles are faces of the 3D tilings \ts2f. 
Consider the set of all these 2D  tilings 
\tsa4 \ perpendicular to the chosen axes. 
The height of the belt allows
(as a necessary condition) for a spacing of
these tilings at Fibonacci distances $S$ and $L$,

\begin{equation}
S=\tau ^3\frac{2}{\tau + 2} \Af \:, \; \; \; \; \; \; \; 
\; \; \;  L= \tau S.
\end{equation}
The same holds true for all 5--fold directions.

A
projection class of tilings by the
Mosseri--Sadoc tiles \cite{D}, \tms  can be locally derived 
from \ts2f \cite{B}. The inflation--deflation class of 
Mosseri--Sadoc tilings\cite{D} has been (not  accurately) 
defined by the substitution 
rules for non--decorated tiles \cite{D}. 
We will come again to this  property.

The third property of the tilings \ts2f \ is that also 
the tilings  \tp \cite{K} can be locally derived 
from \ts2f \cite{A}.  


\section{ The tiles--inflation }

An inflation procedure in a tiling requires two steps:
%
%
\begin{figure}
\vfill 
\caption{ (sektorrep.gif, sektorganz.gif) 
The view on the fundamental sector (lower part)
w.r.t. the icosahedral group of the triacontahedron (upper part).
The fundamental regions marked by all their vertices in 
the sector (lower part) lead to the  coding polytopes 
for all possible 36 vertex configurations in $\EE_\parallel$ 
of the tilings \ts2f .}
\end{figure}
(I) one stretches all the edges of the tiles by the same factor,
(II) one replaces each stretched tile by the tiles of the 
previous size.
Let us call the result the inflated tiling. The rules which 
tell  how one 
replaces each of the tiles are the {\bf inflation rules}. 
In order to find 
the inflation rules for the tilings obtained by the 
projection we 
determine for each stretched representative tile (the tile 
coded by the representative dual in $\EE_\perp$) all its 
possible replacements by the original tiles. The analysis is 
done in the orthogonal space.

When the tiles are stretched in $\EE_\parallel$ by the 
factor $\tau$, the window in $\EE_\perp$, $V_\perp$ and all 
its coding polytopes (the sub--boundaries of the Voronoi domain 
projected to $\EE_\perp$) are transformed by -$\frac{1}{\tau}$,
i.e.  we act by the inflation matrix for the 
module in 6D space by 
$\tau \pp -\frac{1}{\tau} \ps $.
By  construction the inflation
leaves the projection species (the tilings obtained by the 
``cut and project" method) \ts2f \ invariant. 

In order to determine all possible coverings of the stretched 
tiles by the original tiles, it turns out, like in 
a planar tilings \tsa4 \cite{C}
to be enough to 
study how the 36 vertex 
configurations\cite{A,F} go into vertex configurations 
under the inflation. 
A representative vertex configuration (up to  the icosahedral 
symmetry) is a set of tiles 
which fills the solid angle around a vertex point. 
The derivation  of vertex configurations proceeds as follows:
Any tile $X^{i}_\parallel$ in the tilings has a coding polytope
$X^{i}_\perp$ located inside the window
$W_\perp$ of the tiling. $W_\perp$ can be 
decomposed into intersections
$ \bigcap\limits^{l}_{j=1} X^{j}_\perp \neq 0 $.
Any non--empty intersection determines a vertex configuration
$ \bigcup\limits^{l}_{j=1} X^{j}_\parallel $.
One must analyze only the intersections up to the icosahedral 
symmetry. 
For this aim we choose in $\EE_\perp$ a fundamental sector
w.r.t. the icosahedral group from the triacontahedron, see
Figure 3, transform it by $-\frac{1}{\tau}$, obtain the
transformation law of all 36 vertex configurations under the
inflation, and get the following map
\begin{equation}
\begin{tabular}{|c|c|c|c|c|c|cccccccccc|}\hline 
\multicolumn{6}{|c|} {$1$} & \multicolumn{3}{c|} {$2$} & 
\multicolumn{1}{c|}{$3$} & \multicolumn{1}{c|}{$4$} & 
\multicolumn{3}{c|} {$5$} & \multicolumn{1}{c|}{$6$} & 
{$7$} \\ \hline
\multicolumn{2}{|c|} {$\downarrow$} &
\multicolumn{2}{c|} {$\downarrow$} &
\multicolumn{2}{c|} {$\downarrow$} &
{$\downarrow$} & {$\downarrow$} & 
\multicolumn{1}{c|}{$\downarrow$} & 
\multicolumn{1}{c|}{$\downarrow$} & 
\multicolumn{1}{c|}{$\downarrow$} & {$\downarrow$} & 
{$\downarrow$} & 
\multicolumn{1}{c|}{$\downarrow$} & 
\multicolumn{1}{c|}{$\downarrow$} & 
{$\downarrow$}  \\
\multicolumn{2}{|c|} {$8$} & \multicolumn{2}{c|} {$9$} & 
\multicolumn{2}{c|} {$10$} & {$11$} & {$12$} & 
\multicolumn{1}{c|}{$13$} & 
\multicolumn{1}{c|}{$14$} & \multicolumn{1}{c|}{$15$} & 
{$16$} & {$17$} & \multicolumn{1}{c|}{$18$} & 
\multicolumn{1}{c|}{$19$} & {$20$} \\ \hline
{$\downarrow$} &
{$\downarrow$} & 
{$\downarrow$} & 
{$\downarrow$} & 
{$\downarrow$} & 
{$\downarrow$} & 
\multicolumn{1}{c|}{$\downarrow$} & 
\multicolumn{1}{c|}{$\downarrow$} & 
\multicolumn{1}{c|}{$\downarrow$} & 
\multicolumn{1}{c|}{$\downarrow$} & 
\multicolumn{1}{c|}{$\downarrow$} & 
\multicolumn{1}{c|}{$\downarrow$} & 
\multicolumn{1}{c|}{$\downarrow$} & 
\multicolumn{1}{c|}{$\downarrow$} & 
\multicolumn{1}{c|}{$\downarrow$} & 
\multicolumn{1}{c|}{$\downarrow$} \\ \hline
{$23$} & {$24$} & {$22$} & {$25$} & {$21$} & {$26$} & 
\multicolumn{1}{c|}{$27$} & 
\multicolumn{1}{c|}{$28$} & 
\multicolumn{1}{c|}{$29$} & 
\multicolumn{2}{c|} {$30$} & 
\multicolumn{1}{c|}{$31$} & 
\multicolumn{1}{c|}{$32$} & 
\multicolumn{1}{c|}{$33$} & 
\multicolumn{2}{c|} {$34$} \\ \hline
{$\downarrow$} & {$\downarrow$} & {$\downarrow$} & 
$\downarrow$ & $\downarrow$ & $\downarrow$ & 
\multicolumn{10}{c|} {$\downarrow$} \\
{$35$} & {$36$} & {$35$} & {$36$} & {$35$} & {$36$} & 
\multicolumn{10}{c|} {$36$} \\ 
{$\downarrow$} & & {$\downarrow$} & & 
{$\downarrow$} &&&&&&&&&&&\\
{$36$} & & {$36$} & & {$36$} &&&&&&&&&&&\\ \hline
\end{tabular}
\end{equation}

We show that an inflation of the tilings exists. In order to 
define uniquely how each tile of \ts2f \ inflates we find that 
one has to put arrows on each of them such that they break 
its symmetry. One could introduce the arrows on all edges of 
the tiles such that the common edges for the tiles have a 
common orientation of an arrow. Moreover, for two tiles $G$ 
and $C$ we have 
to introduce two different colours, let them be blue and red 
($C^b$, $C^r$, $G^b$ and $G^r$). The substitution matrix $S$ 
for the tiling \ts2f \ is presented as a table
\begin{equation}
\begin{tabular}{c||cccccccc} 
$S$   & $A$ & $B$ & $C^b$ & $C^r$ & $D$ & $F$ & $G^b$ & $G^r$ \\ 
 \hline \hline
$A'$   & $3$ & $0$ & $2$   & $1$   & $0$ & $2$ & $1$   & $2$ \\ 
$B'$   & $0$ & $0$ & $0$   & $1$   & $0$ & $0$ & $0$   & $1$ \\ 
${C^b}'$ & $2$ & $1$ & $0$   & $1$   & $1$ & $2$ & $0$   & $1$ \\ 
${C^r}'$ & $3$ & $0$ & $2$   & $1$   & $1$ & $2$ & $1$   & $2$ \\ 
$D'$   & $0$ & $0$ & $0$   & $1$   & $1$ & $1$ & $0$   & $0$ \\ 
$F'$   & $1$ & $0$ & $1$   & $0$   & $1$ & $1$ & $0$   & $0$ \\ 
${G^b}'$ & $2$ & $1$ & $0$   & $1$   & $0$ & $1$ & $0$   & $1$ \\ 
${G^r}'$ & $3$ & $0$ & $2$   & $1$   & $0$ & $1$ & $1$   & $2$ \\  
\end{tabular}
\end{equation}
and is to be read as for example: $A'$, the tile $A$ stretched 
by $\tau $, is covered by the original tiles $3A$, $2C^b$, 
$1C^r$, $2F$, $1G^b$, $2G^r$. 
Both the arrows on edges that break the symmetry of the tiles
and the different colours are coded in $\EE_\perp$. 
The inflation is not a ``stein--inflation", the 
union of the tiles that cover a stretched tile needs not to 
have the shape of the stretched tile itself. The tiles $B$, 
$F$ and $D$ inflate by a stein--inflation, but not the tiles 
$A$, $C$ and $G$. The faces ${\Sigma}_2$ and ${\Sigma}_3$ 
(golden triangles) inflate as in the tilings \tsa4 \cite{C}. 
The face ${\Sigma}_4$ (small equilateral triangle) transforms 
into ${\Sigma}_1$, $ {\Sigma}_{4}^{'}= {\Sigma}_1 $ (big 
equilateral triangle). The face ${\Sigma}_1$ can't be replaced 
by the union (in the same plane) of other faces. The tiles $A$, 
$C$ and $G$, having each at least one face ${\Sigma}_1$, don't 
have a stein--inflation. In Figure 4 we show the 
stein--inflation of the tiles $B$, $D$ and $F$, 
and in Figures 5 till 8 the inflation of the red tile 
$G^r$, of the tile $A$ and the red tile  $C^r$, of the  
blue  tile $G^b$ and finally of the blue tile $C^b$.

%
%
\begin{figure}
\vspace{8mm}
\vspace{8mm}
\caption{(TauB.gif, TauD.gif, TauF.gif) 
The stein--inflation of the tiles $B$, $D$ and $F$.} 
\end{figure}

%
%
\begin{figure}
\vspace{6mm}
\caption{(TauGr.gif, TauGrback.gif) 
The inflation of the red tile  $G^r$
          seen from two opposite directions.}
\end{figure}

%
%
\begin{figure}
\vspace{8mm}
\caption{(TauA.gif, TauCr.gif) 
The inflation of the tile $A$ and the red tile $C^r$.} 
\end{figure}

%
%
\begin{figure}
\vspace{6mm}
\vspace{6mm}
\caption{(TauGb.gif, TauGbside.gif, TauGbback.gif) 
The inflation of the blue tile  $G^b$ seen from three
          directions.}
\end{figure}

%
%
\begin{figure}
\caption{(TauCb.gif) 
The inflation of the blue tile $C^b$.}
\end{figure}


This inflation rules  define the inflation--deflation
species, the inflation--deflation class of tilings by  six,
or rather eight decorated tetrahedra 
$A, B, C^b, C^r D, F$ and $G^b, G^r$ \cite{Da,GS}.

The question is whether the class of tilings defined by
inflation--deflation (inflation--deflation species \cite{Da})  
is equivalent to the class of tilings \ts2f \ obtained by 
projection (projection species \cite{Da}). From the inflation 
rules for the prototiles of the class of tilings \ts2f \ we 
have derived \cite{E} the inflation rules 
for the prototiles of the projection species \tms \cite{B}.
It turns out that the Mosseri--Sadoc prototiles have also to be
decorated in order to define the inflation procedure 
uniquely \cite{E}.
The tilings by decorated tiles are not only locally
derivable from the projection class \ts2f \cite{A}, 
but equivalent to the class \ts2f. Moreover, the
derived inflation rules for the projection class \tms \ are 
indeed the same as for the inflation species introduced by 
Mosseri and Sadoc (up to a nonexisting decoration)\cite{D,E}.

We determine the volume inflation matrix $M$ for the tiling 
\ts2f :
\begin{equation}
\begin{tabular}{c||cccccccc} 
$M$           & $\{A\}$  & $\{B\}$ &  $\{C^{b}\}$ & $\{C^{r}\}$ & $\{D\}$ 
              & $\{F\}$ & $\{G^{b}\}$  & $\{G^{r}\}$ \\  
    \hline \hline
$\{A'\}$      & {{\small11}$\tau${\small-16}} & 0 & 
                {{\small2}$\tau${\small-2}} & 
                {{\small2}$\tau${\small-3}} &
                0 & {{\small9}$\tau${\small-13}} & 
                {$\tau${\small-1}} & 
                {{\small3}$\tau${\small-4}}\\   
$\{B'\}$      & {\small 0} & {\small 0} & {\small 0} & 
                {\small 1} & {\small 0} & {\small 0} & 
                {\small 0} & {\small 1} \\  
$\{{C^b}'\}$  & {{\small-2}$\tau${\small+4}} & 1 & 0 &  
                {{\small-}$\tau${\small+2}} & 1 & 
                {{\small-}$\tau${\small+3}} & 
                0 & {{\small-}$\tau${\small +2}}\\   
$\{{C^r}'\}$  & {\small-9}$\tau${\small +15} & 0 & 
                {\small-2}$\tau${\small +4} 
              & {\small -}$\tau${\small +2} & {\small 1} & 
                {\small -8}$\tau ${\small +14} 
              & {\small -}$\tau${\small +2} & 
                {\small -2}$\tau${\small +4} \\   
$\{D'\}$      & {\small 0} & {\small 0} & {\small 0} & 
                {\small 1} & {\small 1} & {\small 1} 
              & {\small 0} & {\small 0}  \\   
$\{F'\}$      & {\small 1} & {\small 0} &  {\small 1} & 
                {\small 0} &  {\small 1} 
              & {\small 1} & {\small 0} &  {\small 0} \\   
$\{{G^b}'\}$  & {\small -2}$\tau${\small +4} & {\small 1} & 
                {\small 0} 
              & {\small -}$\tau${\small +2} & {\small 0} 
              & {\small -}$\tau${\small +2} & {\small 0} & 
                {\small -}$\tau${\small +2}\\  
$\{{G^r}'\}$  & {\small -9}$\tau${\small +15} & {\small 0} & 
                {\small -2}$\tau${\small +4} 
              & {\small -}$\tau${\small +2} & {\small 0} & 
                {\small -8}$\tau ${\small +13} 
              & {\small -}$\tau${\small +2} & 
                {\small -2}$\tau${\small +4}.  \\ 
\end{tabular}
\end{equation}

\noindent
In the tabular representation of the matrix $M$ the symbol 
$\{X\}$ stands for Vol$\{X\}$. 
Let us write it as a matrix equation
\begin{equation}
\{X'\}\equiv {\tau}^3 \{X\} = M\{X\}.
\end{equation}
Then the frequencies of the tiles, $f_{X}$ fulfill the equation

\begin{equation}
\tau ^3 f_{X} = M^Tf_{X}.
\end{equation}
So we can test  the inflation rules. The frequencies calculated 
from the volumes of the
corresponding codings in $\EE_\perp$  are

\begin{equation}
\begin{tabular}{c||c|c|c|c|c|c|c|c} 
$X$ & $A$ & $B$ &  $C^{b}$ &  $C^{r}$ &  $D$ & $F$ &  $G^{b}$ &  $G^{r}$ \\ 
 \hline \hline
$f_{X} $ &  $ \frac {1} {2 \tau^0} $  & $ \frac {1} {2 \tau^3} $ 
& $\frac {1} {\tau^2 }$ &  $\frac{1} {\tau^3 } $ & $ \frac {1} { \tau^2} $ 
& $ \frac {1} {\tau } $ & $\frac {1} {2 \tau^3 }$ 
& $\frac {\tau + 2 } {2 \tau^4 } $.\\   
\end{tabular}
\end{equation}
The vector with these entries is indeed an eigenvector of
the matrix $M^T$ with the eigenvalue ${\tau }^3$,
hence fulfills  Equation (10).

One eigenvalue of  the matrix $M$ is $\lambda_0=0$. 
The other eigenvalues fulfill the equation
\begin{equation}
(- \lambda ^{4}+5 \lambda ^{3}-2 \lambda ^{2}-5 \lambda -1 ) 
\times 
[\lambda^3+\lambda^2(13-8\tau)+\lambda(61-38\tau)+62-39\tau]=0
\end{equation}
and are
$\lambda_1=\tau ^3$, 
$\lambda_2=- \tau ^{-3}$,  
$\lambda_3= \tau $, 
$\lambda_4=- \tau ^{-1}$; 
$\lambda_5 = u^+ + u^- $ and\\
$\lambda_{6/7} = - \frac{ u^+ + u^-}{2} \pm i \sqrt {3} 
\frac{u^+ - u^-}{2} $ 
where
$ u^{\pm} = \frac{1}{3\tau } 
\left\{ \left( \frac{75+31\tau}{2} \right) \pm 
\left [ \left( \frac{75+31\tau}{2} \right)^2
-\left( \frac{10}{\tau^2} \right)^3 \right]^{1/2} \right\}^{1/3}
$. Hence $\lambda_5 \approx 1.1868$ and 
$\lambda_{6/7} \approx -0.5934 \pm i 0.7550 $.

\section*{Acknowledgments}
Financial support of Deutsche Forschungsgemeinschaft is 
gratefully acknowledged.
We also thank the Geometry--Center at the
University of Minnesota for making Geomview freely available.

\end{document}